\UseRawInputEncoding
\documentclass[10pt, conference, letterpaper]{IEEEtran}
\IEEEoverridecommandlockouts
\pdfoutput=1
\usepackage{amsfonts, amsmath,amssymb,amsthm}
\usepackage{algorithm, algorithmic}
\usepackage{array}
\usepackage{cases}
\usepackage{changepage}
\usepackage[compress]{cite}
\usepackage{epstopdf}
\usepackage[english]{babel}
\usepackage{etoolbox}
\usepackage{graphicx}
\usepackage{multirow}
\usepackage{physics} 	%	For partial derivative
\usepackage{subfigure}
\usepackage{setspace}
\usepackage{url}  % This makes \url work
\usepackage{lipsum}
\usepackage{nicefrac}	% better inline fraction mode
\usepackage[table,xcdraw]{xcolor}	%Table color
\usepackage{textcomp}% to input temperature unit degrees Celsius
\usepackage{makecell}

\def\ps@headings{%
\def\@oddhead{\mbox{}\scriptsize\rightmark \hfil \thepage}%
\def\@evenhead{\scriptsize\thepage \hfil \leftmark\mbox{}}%
\def\@oddfoot{}%
\def\@evenfoot{}}
\makeatother
\makeatletter
\patchcmd{\@thm}
  {\trivlist}
  {\list{}{\leftmargin=\thm@leftmargin\rightmargin=\thm@rightmargin}}
  {}{}
\patchcmd{\@endtheorem}
  {\endtrivlist}
  {\endlist}
  {}{}
\newlength{\thm@leftmargin}
\newlength{\thm@rightmargin}

\newcommand{\xnewtheorem}[3]{%
  \newenvironment{#3}
    {\thm@leftmargin=#1\relax\thm@rightmargin=#2\relax\begin{#3INNER}}
    {\end{#3INNER}}%
  \newtheorem{#3INNER}%
}

\makeatother

\newtheoremstyle{indentedupright}{3pt}{3pt}{} {}{\bfseries}{.}{.5em}{} % Style 1
\newtheoremstyle{indenteditalic}{3pt}{3pt}{\itshape} {}{\bfseries}{.}{.5em}{} % Style 2

\newcommand{\romu}[1]{\uppercase\expandafter{\romannumeral #1\relax}} % Upper case of roman number, e.g., Rom{1}
\newcommand{\roml}[1]{\lowercase\expandafter{\romannumeral #1\relax}}    % Lower case of roman number, e.g., rom{1} 

\renewcommand{\baselinestretch}{0.99}

\begin{document}
\title{\LARGE Hardware Architecture of Wireless Power Transfer, RFID, and WIPT Systems}
\author{\IEEEauthorblockN{Yu Luo\IEEEauthorrefmark{1}, Lina Pu\IEEEauthorrefmark{2}}
\IEEEauthorblockA{\IEEEauthorrefmark{1}ECE Department, Mississippi State University, Mississippi State, MS, 39759\\
\IEEEauthorrefmark{2}Department of Computer Science, University of Alabama, Tuscaloosa, AL, 35487\\
Email: yu.luo@ece.msstate.edu, lina.pu@ua.edu}
}

\maketitle

%\section{History of WPT and WIT}
%\label{sec:History}
%In this section, we will review the history of WPT and WIT technologies separately. After that, we introduce the development of WIPT technology. 

%=============================================
%=============================================
\section{Historical Milestones of Wireless Power Transfer (WPT)}
\label{subsec:HisWPT}
The early effort on WPT can be traced back to late 1950s when a theoretical analysis from Goubau and Schwering showed that power could be transmitted over any distance with near $100\%$ efficiency through a concentrated beam~\cite{goubau1961guided}. Three years later, this theory was confirmed by an experimental demonstration indicating the born of an effective WPT system~\cite{degenford1964reflecting}. Around 1963, the rectenna was invented, which is a memorable event in the history of the WPT development. The rectenna can efficiently convert electromagnetic (EM) energy arrived at a WPT receiver into the direct current (DC). The invention of the rectenna paved the way of long-distance WPT.

In the early stage of WPT development, a representative experiment was to power an unmanned helicopter with the microwave energy reflected from an ellipsoidal reflector placed at the ground~\cite{george1963efficient} . The small helicopter was able to hover several meters above the reflector without extra energy supply~\cite{brown1966experiments}. With an increased energy conversion efficiency in  $2.4$\,--\,$2.5$ GHz frequency bands, the flight altitude of the microwave-powered aircraft was further improved. In 1988, Canadian Stationary High Altitude Relay Platform Program (SHARP) demonstrated an airplane with $4.57$\,m wingspan. The airplane could fly at $50$\,m altitude for $3.5$ minutes through receiving energy radiated from a parabolic dish with $10$\,kW transmission power~\cite{schlesak1988microwave}.

In 1968, Peter Glaser introduced a concept of solar power satellite (SPS) that captures solar power through the satellite running in the geostationary orbit, and then sends the harvested energy back to the earth via microwaves~\cite{glaser1968power}. To examine the feasibility of a long range WPT from a satellite to the ground, a successful experiment was conducted by Raytheon company in 1975. Between 1978 and 1986, DOE (Department of Energy) and NASA (National Aeronautics And Space Administration) jointly investigated the satellite power system concept development and evaluation program~\cite{dietz1981satellite}. Nowadays, SPS is still considered as one of potential candidates to replace the fossil fuel and nuclear energy for a green, safe, and sustainable power supply~\cite{sasaki2013microwave}.

%=============================================
%=============================================
\section{Historical Milestones of Wireless Information Transfer (WIT)}
\label{subsec:HisWIT}
Since Maxwell's equations was published and later verified by Heinrich Hertz in 1888, the wireless communications has well prepared to walk into the real world. In 1902 and 1914, the frequency modulation (FM) and the amplitude modulation (AM) was proposed respectively for wireless communications. Until today, those two modulation schemes are still widely used for the radio broadcasting. 

1948 is a year worth remembering for us. In this year, Claude E. Shannon laid out the foundation of information theory in his paper entitled ``A mathematical theory of communication''~\cite{shannon1948mathematical}. In this paper, Shannon for the first time defined the channel capacity and described it with a mathematical model. Thereafter, researchers have a mathematical model to accurately calculate the fundamental limits on the wireless communication channel.

In 1979, the first generation (1G) of wireless cellular technology was launched in Japan. It applied the analog telecommunication technology, and then replaced by the second generation (2G) of digital telecommunication technology. In late 1991, GSM (global system for mobile communications), which was the world first TDMA (time-division multiple access) standard, was developed for 2G networks~\cite{huurdeman2003worldwide}.
 
Today, we are experiencing the high speed fourth-generation long term evolution (4G LTE)  broadband cellular network, and expecting for more surprise, like the virtual reality (VR) and the edge computing, that will be brought by the fifth-generation (5G) mobile network in the near future. Compared with the 2G GPRS (general packet radio service), which provides up to $114$\,kbps of download and $20$\,kbps of upload data rates, the 4G network supports up to $100$\,Mbps and $1$\,Gbps data rates for applications with high mobility and low mobility, respectively~\cite{milos20142G}. These two rates will turn out to be at least ten times higher in the incoming 5G communications~\cite{andrews2014will}.

%=============================================
%=============================================
\section{History of  Wireless Information and Power Transfer (WIPT)}
\label{sec:HisWPT}
The original developments of WPT systems and WIT systems are independent. The former had a huge size in order to achieve an efficient energy transfer at a high-power level~\cite{brown1984history}, while the latter aimed to reduce the device size for high-speed communications with a low power consumption. Until 1948, Harry Stockman introduced a new concept, called the communication by means of reflected power~\cite{stockman1948communication}. His work is the prototype of the radio frequency identification (RFID), which is the first attempt to combine WIT with WPT.

Three decades later, a short-range RFID system was implemented  based on the modulated backscatter~\cite{koelle1975short}. In this system, the RF reader sent a $4$\,W of continuous wave (CW) at $1$\,GHz to an RF tag, which then changed the load on its rectifier to modulate the amplitude of backscatter waves for the data transmission. Since a portion of CW energy from the RF reader was rectified to power the RF tag for wireless communications, the passive RFID, can be considered as the first WIPT system in the history. Nowadays, RFID and its derivatives, near-field communication (NFC)~\cite{want2011near}, have been widely used in different places for item classification, electronic payment, and tool management~\cite{landt2005history, singh2006state}.

\begin{figure}[htb]
\centerline{\includegraphics[width=8.5cm]{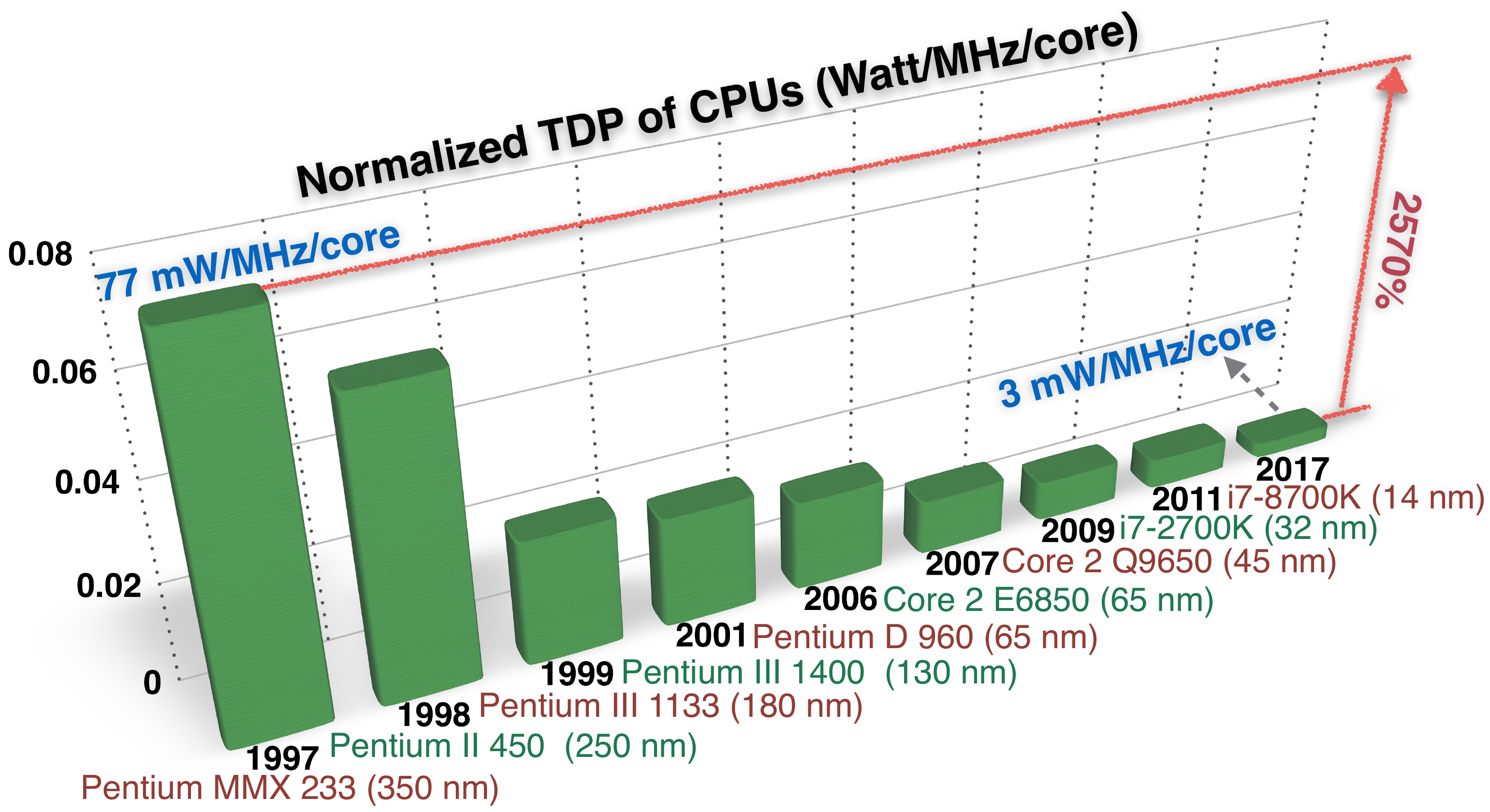}}
  \caption{With the progress of semiconductor technology, the normalized  TDP of Intel's processors decreasesk significantly.}\label{fig:cpuPow}
\end{figure}

The invention of integrated circuit (IC) further promoted the integration of WPT and WIT~\cite{Nobel2000Nobel}. With the rapid development of the IC technique, the computational speed and the power consumption of semiconductor chips had a drastic improvement~\cite{koomey2011implications}. Taking Intel's microprocessors as an example, the normalized thermal design power (TDP)\footnote{We calculate the Normalized TDP\,=\,$\displaystyle\frac{\text{TDP of processor}}{\text{Clock rate}\times\text{Number of cores}}$.} of Pentium MMX 233 released in 1997 was $77$\,mW/MHz/core~\cite{Cpu2017Cpu}. In 2017, this value was reduced to $3$\,mW/MHz/core in Core i7-8700K, only $3.9\%$ of Pentium MMX 233~\cite{Intel2017Intel}, as shown in Fig.~\!\ref{fig:cpuPow}.

\begin{figure}[htb]
\centerline{\includegraphics[width=6.0cm]{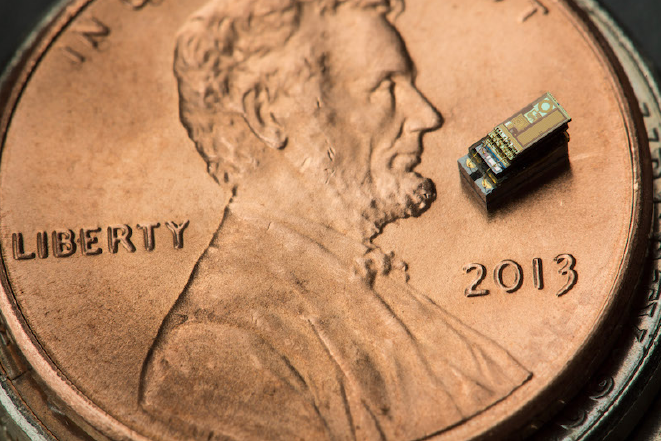}}
  \caption{A smart dust, Michigan Micro Mote (M$^3$) developed by University of Michigan, sitting on a penny~\cite{michigan2015michigan}.}\label{fig:smartDust}
\end{figure}

The low TDP of modern microprocessors makes them realistic to perform some complicated tasks, e.g., fast Fourier transform (FFT), coding, and decoding with an extremely low energy consumption. In 1990s, the concept of smart dust was proposed. As demonstrated in Fig.~\!\ref{fig:smartDust}, the smart dust is a miniature wireless mote. It is constructed as an aggregation of multiple tiny microelectromechanical systems (MEMS) for environment sensing and wireless communications~\cite{cook2006soc}. Different from the passive RFID that obtains energy from a dedicated RF reader, the smart dust can harvest energy from surrounding environment, such as solar, vibrations, or even ambient RF waves~\cite{liu2013ambient}, for a perpetual operation. Owing to the features of miniaturization and  sustainability, the smart dust is gradually penetrating into our daily life~\cite{dorrier2017smart, aquila2017smart}. 

In recent years, the idea of WIPT is attracting people's attention~\cite{varshney2008transporting}. In WIPT, an RF facility can transmit  energy and information to an associated wireless devices at the same time. The scenarios with both a co-located and a separate energy and information receivers were explored in \cite{zhang2013mimo} for WIPT running in a multiple-input and multiple-output (MIMO) network. Today, researchers are putting efforts on integrating WIPT system into the 5G massive MIMO communications. They aim to transmit the  energy flow and the data flow to  wireless devices flexibly through narrow beams for efficient and sustainable wireless communications~\cite{wu2017overview, liu2015integrated}.

\section{Insight into WPT}
\label{sec:Insight}
This section focuses on the WPT technology, which is the foundation of WIPT. We first introduce the features of EM waves, and then discuss several common WPT methods that can combine with the WIT technology for WIPT.

%============================================
%============================================
\subsection{Features of EM Waves}
\label{subsec:FeaEM}
The EM wave is the carrier of both energy and information in WIPT systems. Its behavior depends upon the propagation distance, the wavelength, and the size of the transmitting antenna. In general, an alternating EM field can be divided into following three regions~\cite{constantine2015antenna, stutzman2012antenna}:
\vspace{0.1cm}
\begin{adjustwidth}{-0.77cm}{0cm}
\begin{description}
\setlength{\labelsep}{-0.95em}
\itemsep 0.05cm
  \item[a)] \emph{Reactive near field (induction region)}. The boundary of this region is $R<0.62\sqrt{D^3/\lambda}$, where $R$, $D$, and $\lambda$ are the propagation distance, the dimension of the transmitting antenna, and the EM wavelength, respectively. The induction region has two critical features. First, the electric and magnetic fields are separate. Second, if there is no receiving device, the EM energy flows around the transmitting antenna without any loss~\cite{umenei2011understanding}.
   \item[b)] \emph{Radiating near field (Fresnel region)}. It is a region bountied by $0.62\sqrt{D^3/\lambda}<R<2D^2/\lambda$. The induction and  radiating fields coexist in the Fresnel region. Different from induction field, the energy in the radiating field cannot be retrieved once it leaves the transmitting antenna. Moreover, the radiation pattern of an EM field in this region varies with the propagation distance, hence the relationship between the electric field and the magnetic field is complex in the Fresnel region.
  \item[c)] \emph{Far field (Fraunhofer region)}. When the propagation distance of an EM wave is longer than $2D^2/\lambda$, it enters the far-field region. In this region, the radiating field is the dominator, which determines the antenna's transmission pattern. The electric field and the magnetic field in the far field cannot be separated anymore but propagate together as a wave. In the Fraunhofer region,  the electric field, the magnetic field, and the propagation direction of EM waves are always orthogonal to each other. The radiation pattern of EM waves in the far field does not change with the distance. 
\end{description}
\end{adjustwidth}
\vspace{0.1cm}

\begin{figure}[htb]
\centerline{\includegraphics[width=8.5cm]{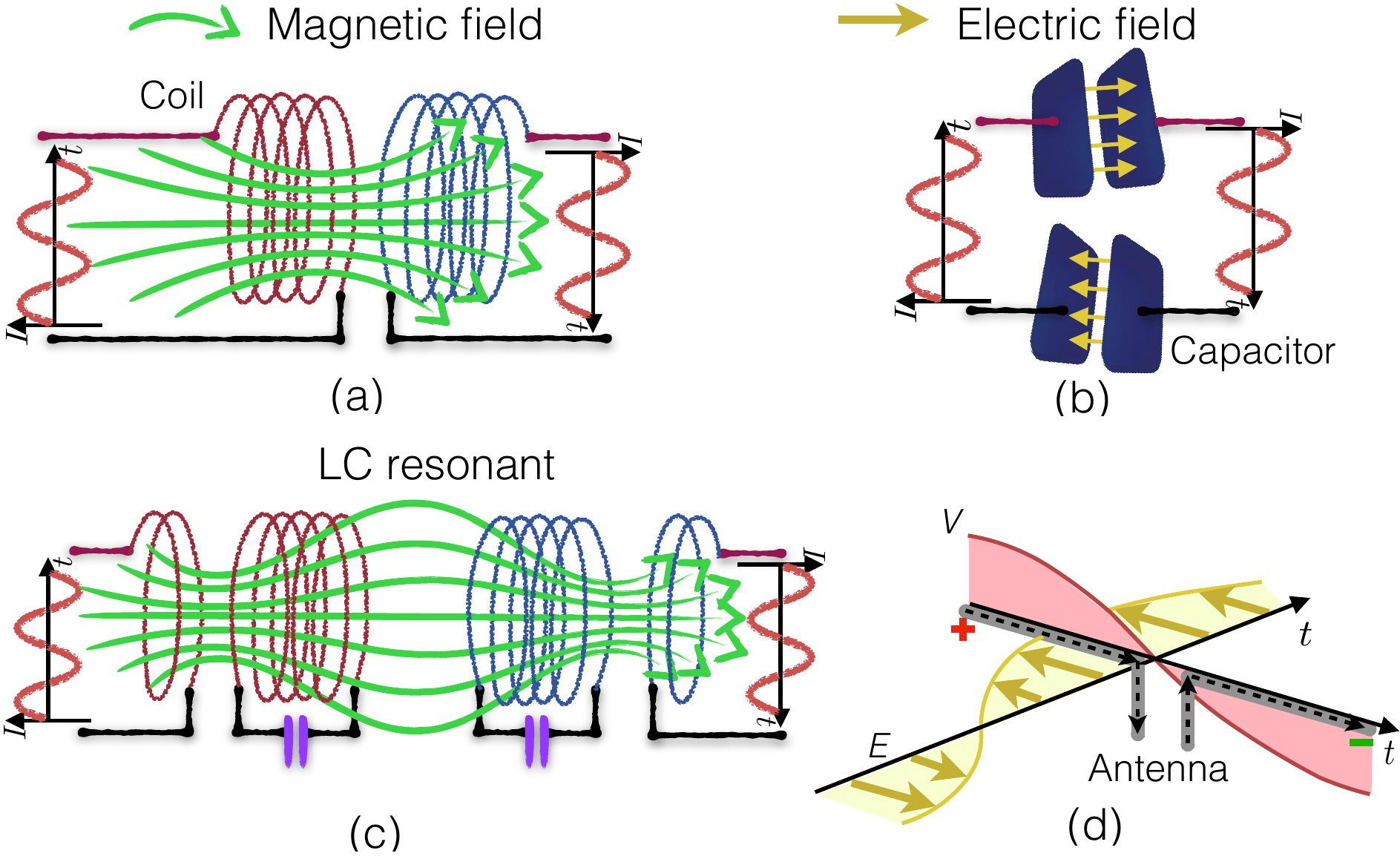}}
  \caption{Four different WPT methods, where $I$, $V$, $t$, and $E$ represent the current, the voltage, the time, and the electric field, respectively. (a) Near-field inductive coupling WPT. (b) Near-field capacitive coupling WPT. (c) Middle-range resonant inductive coupling WPT. (d) Far-field radiative WPT.}\label{fig:WPT}
\end{figure}

By utilizing the properties of EM waves in above three regions, people developed four representative WPT methods: the near-field inductive coupling WPT, the near-field capacitive coupling WPT, the middle-range resonant inductive coupling WPT, and the far-field radiative WPT, as depicted in Fig.~\!\ref{fig:WPT}. In following sections, we give a briefly introduction on each of the four methods.

%============================================
%============================================
\subsection{Near-Field Inductive Coupling WPT}
\label{subsec:NeaInductive}
In the near-field inductive coupling WPT system, antennas at the sender and the receiver sides are two subtending coils to transfer energy wirelessly through the magnetic field, as demonstrated in Fig.~\!\ref{fig:WPT}(a). To work properly, the transmitting coil needs to connect with an alternating current (AC) power supply to generate an oscillating magnetic field. According to the Faraday's law,  AC is generated at the receiver side once the oscillating magnetic field passes through the receiving coil~\cite{sadiku2014elements}.

The inductive coupling WPT can only work efficiently in the reactive near field region. Therefore, the transmit-receive coils must be close to each other so that most of the induction lines produced by the transmitter can extend to the receiving coil. Furthermore, the dimension of the transmitting coil needs to be much smaller than the wavelength of the EM field so the energy cannot escape from the radiation of EM waves. Therefore, in an inductive coupling WPT system, the frequency of the AC power supply should be low and the wavelength of the EM field should be long~\cite{hall1980the, smith1997introduction}. 

The inductive coupling WPT technology has been widely used in the application of wireless charging. For instance, the electric toothbrush can directly use the $50/60$\,Hz AC from an outlet to charge the battery; a smartphone like iPhone X and Samsung Galaxy S8 that supports the inductive mode of the Qi standard\footnote{Recent Qi receivers can be charged in both inductive mode and the resonant mode~\cite{wireless2017a}. The inductive mode can provide a high power efficiency, while the resonant mode can increase the charging distance.} can be changed at several kilohertz frequency~\cite{poole2017qi}. 

\begin{figure}[htb]
\centerline{\includegraphics[width=8.5cm]{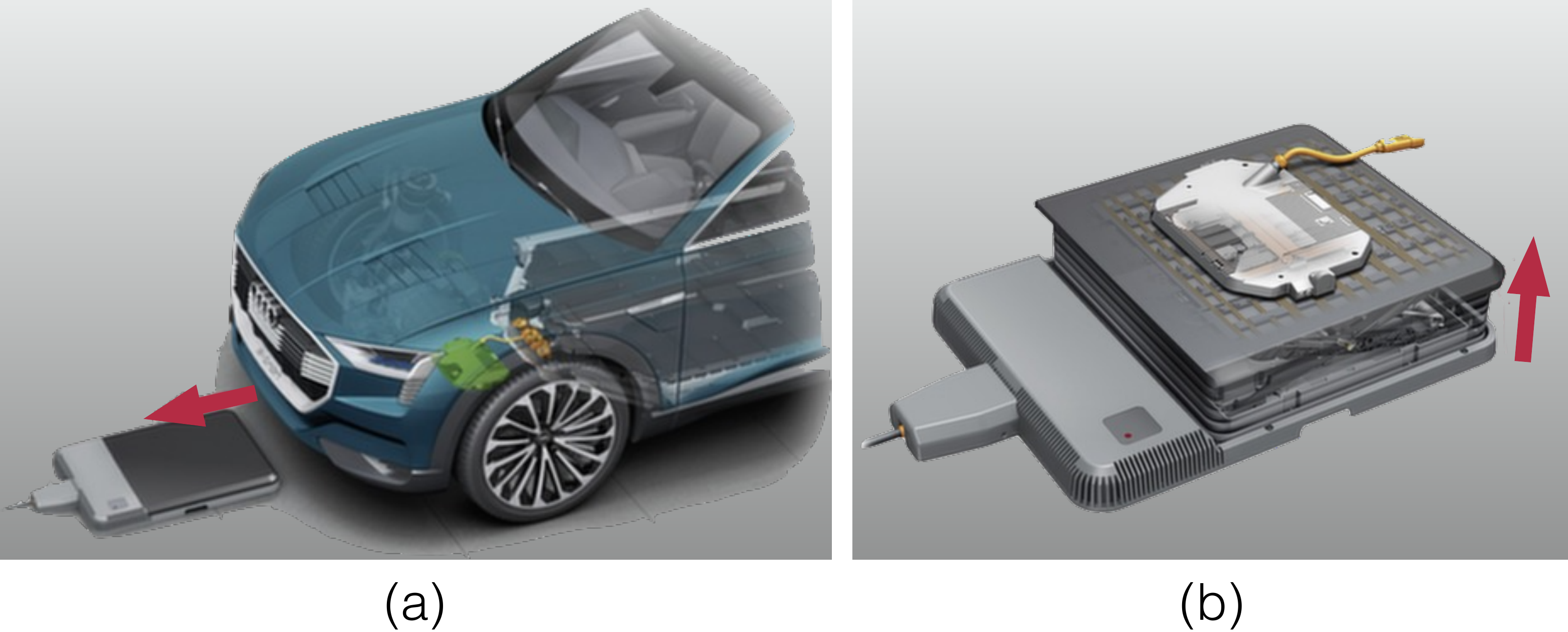}}
  \caption{The Audi wireless charging system (recreated from \cite{audi2015audi}). (a) Move a vehicle to the charging place. (b) Raise the floor plate to change the vehicle wirelessly.}\label{fig:AudiWPT}
\end{figure}

Today, the inductive coupling WPT has become an alternative to charge  electric vehicles~\cite{bombardier2016change, lukic2013cutting}. To achieve this, automobile manufacturers integrate the receiving coil into a vehicle's chassis and the transmitting coil is placed at an independent floor plate, as demonstrated in Fig.~\!\ref{fig:AudiWPT}. As announced in \cite{audi2015audi}, through using the inductive AC coupling technology developed by Audi automobile manufacturer in 2015, the wireless charging system can offer a power of $3.6$\,kW, and a higher power up to $11$\,kW is possible in a future version.

%============================================
%============================================
\subsection{Near-Field Capacitive Coupling WPT}
\label{subsec:Neacapacitive}
Different from the inductive coupling method, which utilizes the magnetic field for power transfer, the energy in a capacitive coupling WPT system is transmitted by an alternating electric field, as shown in Fig.~\!\ref{fig:WPT}(b). In such a system, a pair of conductive plates are placed closely to form  electrodes of a capacitor. When an AC power supply is connected to the transmitting plate, an alternating electric field is created. It causes an oscillating electric potential on the receiver plate through the electrostatic induction and then generates AC at the receiver side.

In order to transmit energy efficiently, a high-voltage electric field needs to be created between the electrodes of a capacitor. However, through dielectric polarization, an electric field can interact with dielectric materials, such as the air and tissues, that are poor conductors of electricity but efficient supporters of the electrostatic field~\cite{gabriel1996dielectric2}. 
A strong electric field can also affect human body directly or produce noxious ozone to damage lungs
\cite{eliasson1987ozone, devlin1991exposure, gabriel1996dielectric1}. 
Therefore, the capacitive coupling method has much less commercial applications than the inductive coupling approach for WPT.

%============================================%============================================
\subsection{Middle-Range Resonant Inductive Coupling WPT}
\label{subsec:MidResonant}

Resonant inductive modified the inductive coupling method to extend the distance of power transmission. In a resonant inductive coupling system, an additional LC resonance circuit that consists of a coil and a capacitor is added at the receiver side. If the frequency of the EM field created by the sender's coil is equivalent to the resonant frequency of LC circuit, the phase of the magnetic field is synchronized between the transmitting and receiving coils, which can significantly improve the energy transfer efficiency. The resonant inductive coupling method can work at medium-range, which is defined as somewhere between one and ten times of the diameter of the transmitting coil~\cite{agbinya2015wireless}.

To further extend the distance of power transmission, MIT developed a new technique, called WiTricity, in 2007. In WiTricity, both the sender and receiver sides have LC circuits with the same resonant frequency, as illustrated in Fig.~\!\ref{fig:WPT}(c). As reported in \cite{kurs2007wireless}, the energy efficiency of WiTricity is around $40\%$ when transmitting $60$\,W of power at $9.9$ MHz over the distance up to $6.6$\,ft (8 times of the coil's radius).

\begin{figure}[htb]
\centerline{\includegraphics[width=8.7cm]{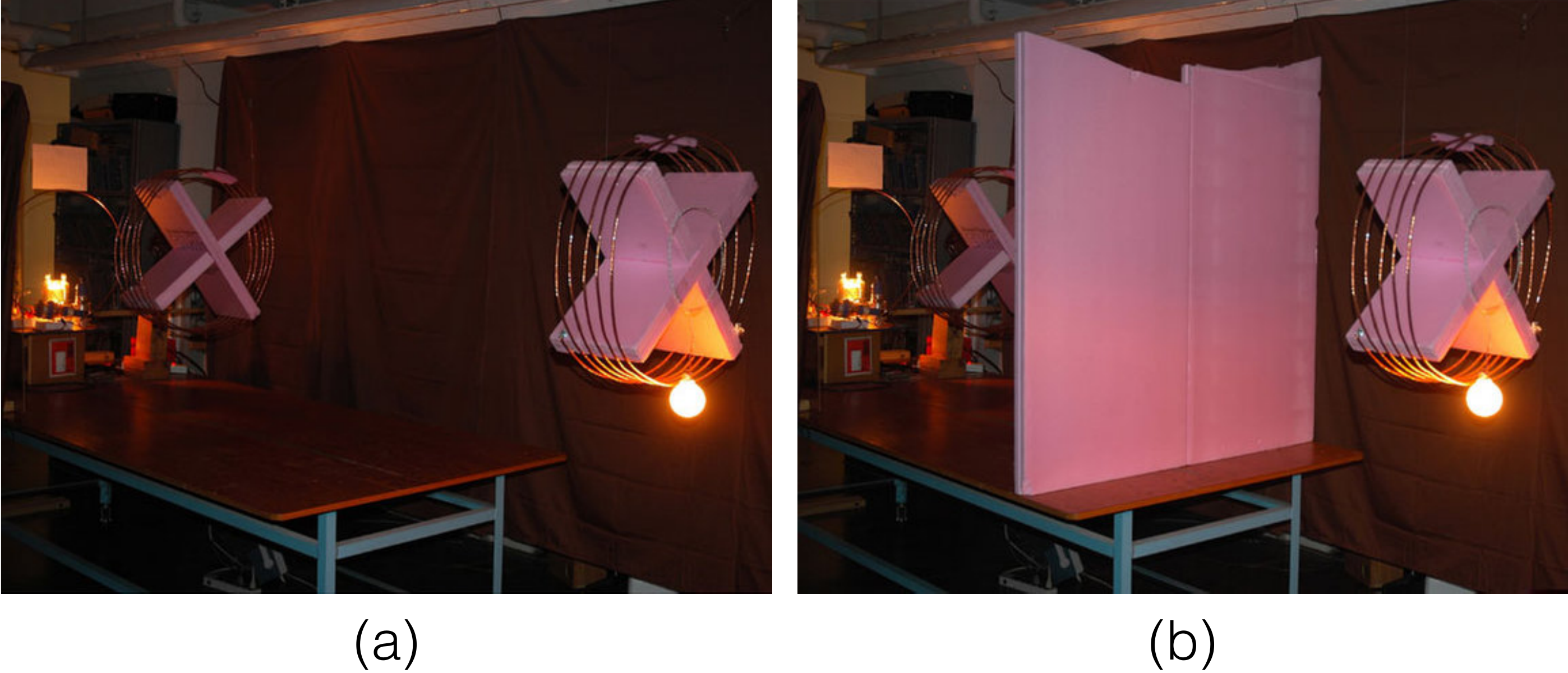}}
  \caption{Lighting a $60$\,W bulb 6.6 feet away from an energy source through the WiTricity technology\cite{choi2007wireless}. (a) Without obstruction. (b) A wooden panel is placed in the middle to block the direct path of the EM field.}\label{fig:}
\end{figure}

Today, the resonant inductive coupling technology has been widely applied to charge mobile devices that support the resonant mode of the Qi standard~\cite{wireless2017a}. It is also a promising technology to enable an on-line WPT system~\cite{lee2010line}, in which the transmitting coil is designed as a continuous power line laid underground or multiple distributed power stations on the roadway. The power transfer initiates once a vehicle runs along the power line or gets close to the power station~\cite{miller2014demonstrating, musavi2012wireless}. By using the on-line WPT technique, the battery capacity on the vehicle can be reduced by $20\%$ compared to conventional electric cars~\cite{ahn2011magnetic}. This can greatly reduce the cost, weight, and size of vehicles.

In 2013, Korea Advanced Institute of Science and Technology (KAISA) developed the world'€™s first WPT electric bus network~\cite{barry2013in}, which consists of 15 miles of road in Gumi, South Korea. To power the bus, the transmitting coil is buried 12 inches below the road surface, and bus chassis is 6.7 inches from the ground. The total length of the power line takes up between 5 and 15 percent of the entire route. The frequency of the EM field used for WPT is $20$\,kHz, which can charge the bus at $100$\,kW with $85\%$ efficiency~\cite{anthony2013world}.

%============================================%============================================
\subsection{Far-Field Radiative WPT}
\label{subsec:FarRaditative}

For the far-field WPT technology, the transmission range of EM energy can be much longer than the size of the transmitter. In generally, far-field WPT systems can be divided into two categories: high-power applications and  low-power applications. The former have been introduced in Section~\!\ref{subsec:HisWPT}, which includes SPS and the high-density microwaves powered aircraft; the latter will be discussed latter.

A high-power WPT system usually needs a parabolic reflector to form a narrow beam, which can significantly reduce the spreading loss of EM waves in the free space. However, a narrow beamformer needs the reflector's aperture to be much larger than the wavelength. Using SPS that studied by NASA in 1979 as an example, if a satellite collects energy in the space and then transmit it back to the earth through a $2.4$\,GHz microwave, the diameters of transmitting and the receiving reflectors need to be $1$\,km and $10.5$\,km, respectively~\cite{landis2006re}. Apparently, it is infeasible to use such a huge high-power WPT system in a far-field WIPT device that commonly having a portable size.

In a low-power far-field WPT system, the energy is  radiated in the form of EM waves through an antenna, as demonstrated in Fig.~\!\ref{fig:WPT}(d). To maximize the energy transfer efficiency, the antenna needs to work in the resonant state. This requires the length of the receiving antenna no less than half wavelength of EM signals so that a stationary wave\footnote{Stationary wave, also called standing wave, is a combination of two waves moving in opposite directions, each having the same amplitude and frequency. Due to the constructive interference, the amplitude of a stationary wave is the sum of amplitudes of the two waves.} with the maximum current is created in the antenna~\cite{hall1980the}. Compared with a high-power WPT system, the size of a low-power far-field WPT system can be small, which makes it fit the WIPT applications very well.

\section{Architecture of WIPT System}
\label{sec:Architecture}

This section studies the implementation of WIPT, which combines the WPT methods discussed in Section~\!\ref{sec:Insight} with WIT technique. In order to better understand the working principle of WIPT, we introduce the architectures of three modern WIPT systems, they are the near-field inductively coupled WIPT, the backscatter WIPT, and the far-field WIPT.

%=============================================
\subsection{Near-Field Inductively Coupled WIPT}
\label{sec:ArchInductive}

The inductively coupled WIPT is widely used in RFID and NFC systems~\cite{weinstein2005rfid, coskun2015survey}, where the data is delivered in a passive manner. Using RFID as an example, through the inductive coupling WPT, an RF tag can receive energy from an alternating EM field created by an associated reader. Due to the mutual inductance, an imaginary impedance will present at the reader side once the tag gets close to it. The voltage at the reader's antenna (coil) is proportional to the imaginary impedance, which decreases with the tag's load resistance~\cite{finkenzeller2010rfid}. Therefore, if the tag can modulate its load resistance based on the data to be transmitted, the reader can receive that data by measuring the voltage variation at its antenna.

\begin{figure}[htb]
\centerline{\includegraphics[width=8.5cm]{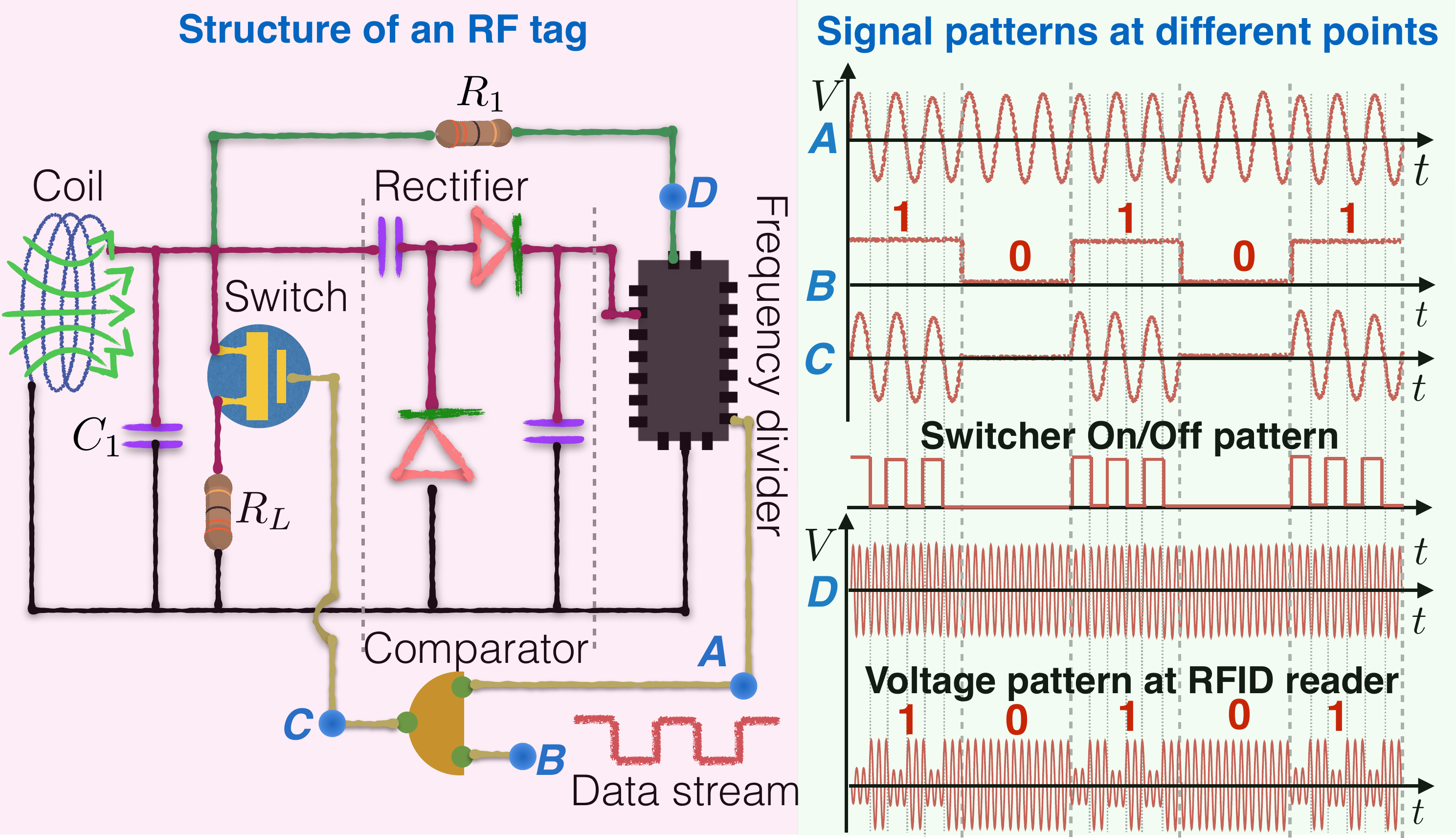}}
  \caption{An example of a near-field inductively coupled WIPT system. The left part is the hardware design of the tag and the right part indicates signal patterns collected at different positions of the system.}\label{fig:structure(a)}
\end{figure}

Fig.~\!\ref{fig:structure(a)} gives an example of how an inductively coupled RF tag sends data through the load modulation. As shown in the left part (pink frame) of the figure, a capacitor $C_1$ is connected with the transmitting coil in parallel to form an LC circuit. The resonant frequency of the LC circuit is equivalent to the frequency of alternating EM field generated by the reader. When the EM signal is received by the LC circuit, it is divided into two parts. The first part is converted into energy to power the tag. The second part goes through a protective resistance $R_1$ to provide a frequency divider a basis signal at point D of Fig.~\!\ref{fig:structure(a)}. After going through the frequency divider, a low-frequency subcarrier is available at point A. Afterward, the subcarrier signal and the data stream enter a comparator to generate a modulated signal. Specifically, only if the data bit is ``1'' (i.e., high voltage), the subcarrier signal can pass through the comparator; otherwise, the comparator's output is zero, as illustrated at point C. 

To achieve load modulation, the comparator is connected to a field-effect transistor (FET) as a switch for an on-off control. The switch is turned ON/OFF if the output of the comparator is a positive/nonpositive value; the on-off pattern of the switch is depicted in the right part (green frame) of Fig.~\!\ref{fig:structure(a)}. Once the switch is turned on, a resistance $R_L$ is connected to the transmitting coil in parallel. This reduces the load resistance of the RF tag, thereby decreasing the voltage at the reader's antenna, as shown in the figure.

An inductively coupled WIPT system needs to work in the induction region of an EM field; therefore, its communication range is short, usually within $1$\,m. Moreover, as discussed in Section~\ref{subsec:NeaInductive}, to transmit energy efficiently, the wavelength of EM field needs to be much larger than the size of the transmitting coil. Therefore, the frequency of the basis signal sent from a reader should be low, which is commonly below $135$\, kHz and $13.56$\,MHz for low-frequency and high-frequency applications, respectively~\cite{international2009iso, international2010iso}. As a consequence, the data rate of an inductive coupled WIPT system is low. For example, the data rate of NFC and RFID can only reach $848$\,kbps\footnote{Depending on applied coding schemes, the transmission speed of NFC has four different options: $106$, $212$, $424$, or $848$\,kbps~\cite{nxp2013mifare}.} when the basis signal of the reader is  $13.56$\,MHz.

%=============================================
\subsection{Backscatter WIPT}
\label{sec:ArchBack}

In a backscatter WIPT system, a reader radiates a carrier signal to an RF device, which can then modulate the intensity of a reflected carrier signal in accordance with the data stream to be transmitted. Thereafter, the reader can detect the amplitude fluctuation of reflected waves to decode the data. Next, we introduce two different backscatter WIPT techniques.

%=======================================
\subsubsection{Conventional backscatter WIPT}
The idea of conventional backscatter WIPT is to adjust the strength of scattered signals. This can be achieved by changing the radar cross-section (RCS) of an antenna ~\cite{knott2012radar}. In backscatter WIPT systems, a high RCS indicates that given the incident power of EM waves, an antenna can generate a strong reflected wave and vice versa. The total RCS of an antenna can be divided into two components, the structural mode RCS and the antenna mode RCS. The former is determined by the antenna's physical feature (e.g., shape, structure, and material); the latter is caused by the antenna's reciprocity, which radiates reflection signals as a transmitter depending on the operating status of an antenna (e.g., open circuit, short circuit, or resonance). Specifically, if the impedance of a receiving antenna doesn't match the circuit's load, the incident wave will be partly radiated back into the air~\cite{stutzman2012antenna}. Eventually, the structural scattering field and the antenna scattering field superimpose to form the total scattered field.

\begin{figure}[htb]
\centerline{\includegraphics[width=8.5cm]{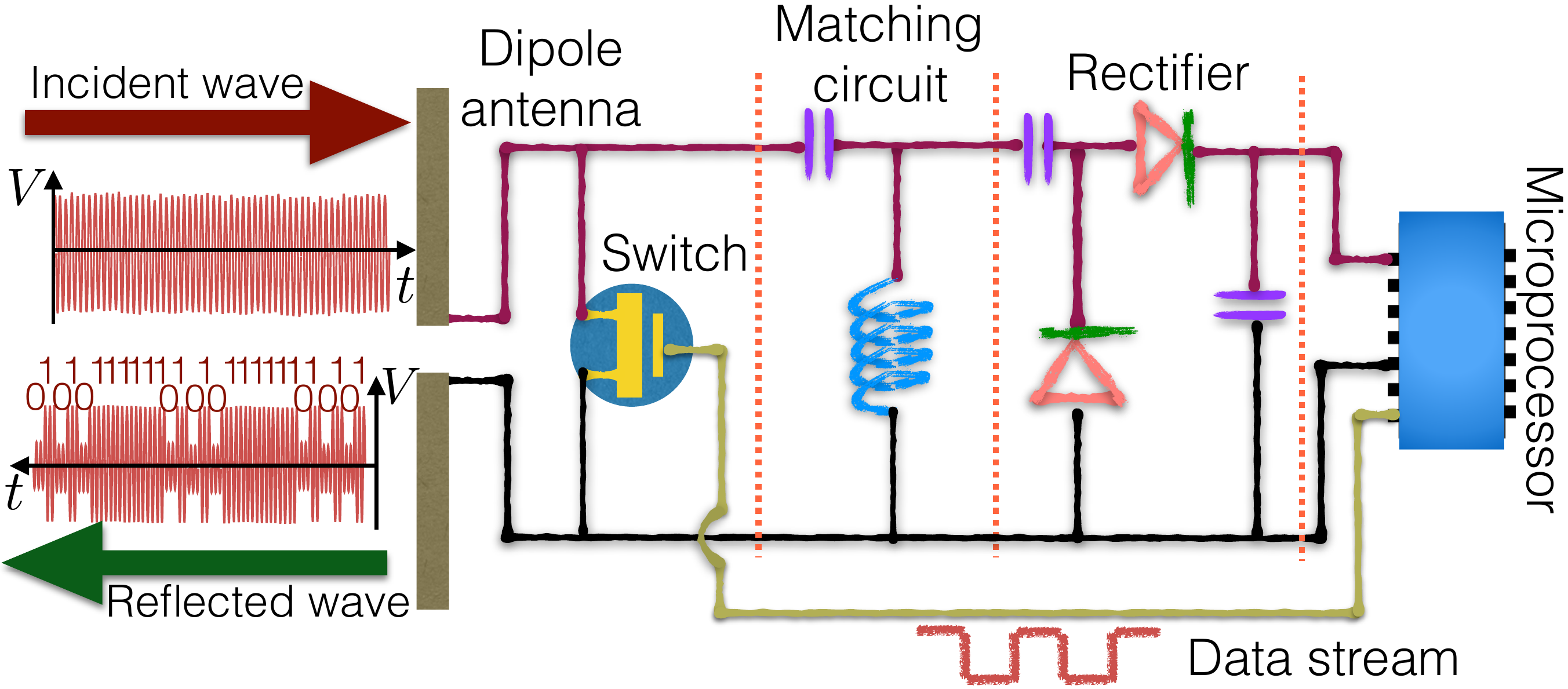}}
  \caption{An example of a backscatter WIPT system, which modulates the reflected signal by switching the antenna's status between the short circuit and the resonance for a data transmission.}\label{fig:structure(b)}
\end{figure}

The RCS is maximized when an antenna is short circuited; it is minimized when the antenna's impedance well matches the circuit's load~\cite{knott2012radar}. Therefore, a backscatter WIPT system can modulate the amplitude of reflected waves by changing the status of its receiving antenna for the data transmission. In Fig.~\!\ref{fig:structure(b)}, we provide an example of a backscatter WIPT system. In the figure, an FET switch is connected with a dipole antenna in parallel. The ON/OFF of the switch is manipulated by the data sequence to be transmitted. When the switch is ON, the antenna is short-circuited to maximize RCS, thereby causing a strong  backscattered wave. When the switch is OFF, the matching circuit makes the impedances of the antenna and the circuit load identical to minimize the backscattering strength. Through measuring the amplitude of reflected waves, the reader can decode the information from the sender. 

\begin{figure}[htb]
\centerline{\includegraphics[width=6.5cm]{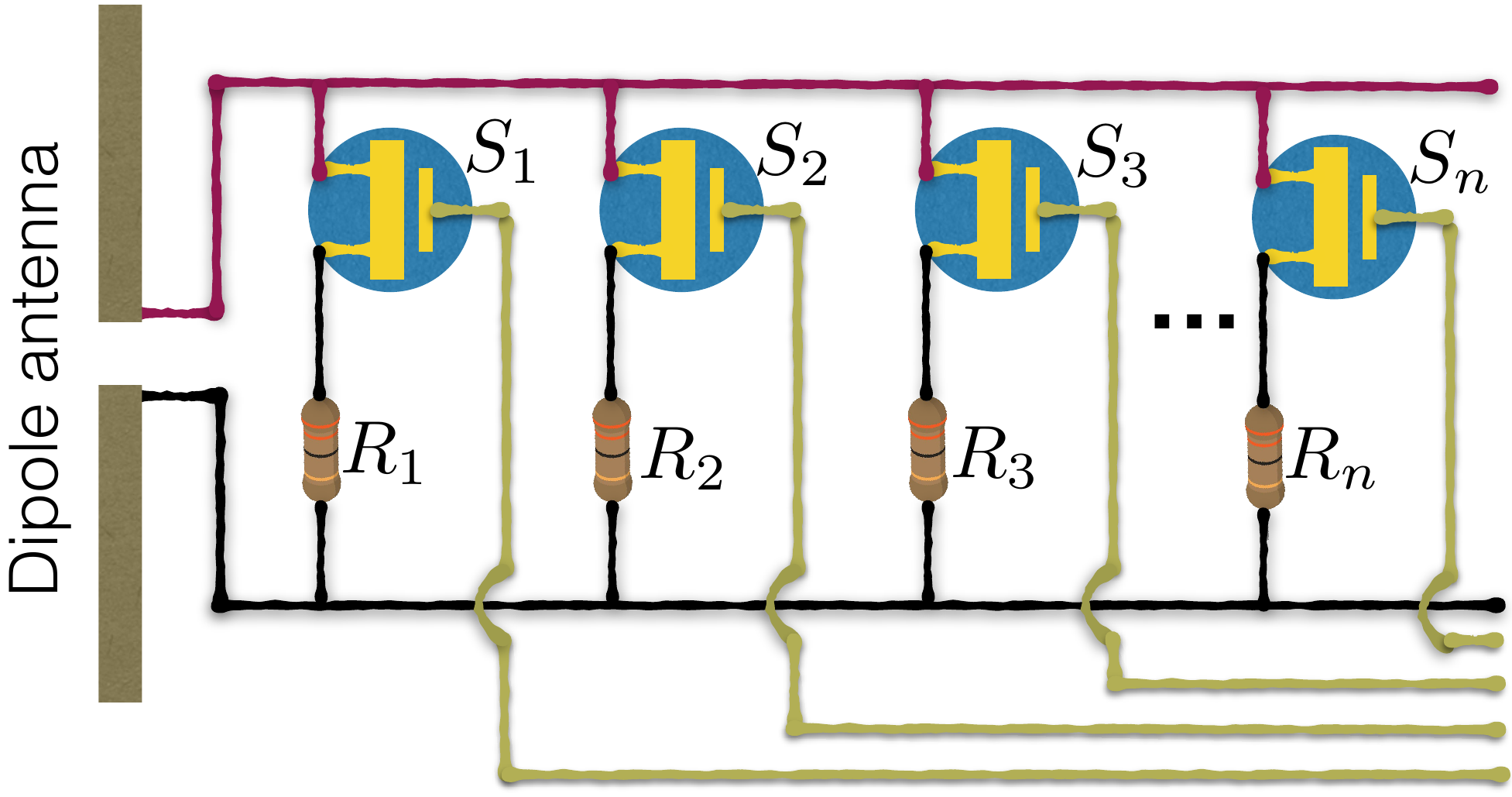}}
  \caption{Frontend of a backscatter WIPT circuit to generate reflected QAM signals, where $S_i$ and $R_i$ represent the $i^{th}$ FET switch and the $i^{th}$ resistor, respectively.}\label{fig:structure(c)}
\end{figure}

To increase the data rate of a backscatter WIPT system, we can adopt a quadrature amplitude modulation (QAM) at the sender side~\cite{thomas2010qam}. This can be implemented by connecting the transmitting antenna with multiple different resistors in parallel, as shown in Fig.~\!\ref{fig:structure(c)}. During communication, $L$-bits data sequence turns on one out of $n$ switches and then the connected resistor creates a certain level of load mismatch, where $L\!=\!log_2(n)$. Accordingly, the reflected waves can have $n$ different strengths, and each strength can represent $L$-bits data.

The communication range and the data rate of a conventional backscatter WIPT system are mainly determined by the radiation power of the reader and the size of the transmitting antenna. Owing to the high carrier frequency, the transmission rate of a backscatter WIPT system can reach $2$\,Mbps or even higher, much faster than that in an inductively coupled WIPT system~\cite{thomas2013modulated}. The communication rage of a conventional backscatter WIPT system is usually within $12$\,m due to a high spreading loss of scattering signals and an imperfect reflection of the transmitting antenna~\cite{international2013iso}. As measured in \cite{penttila2006radar}, if a reader uses $1$\,mW power to transmit a $900$\,MHz carrier signal via a $6$\,dBi antenna and an RF tag is placed $2$\,m away from the reader to reflect the carrier signal through a $100$\,mm$\times$$100$\, mm patch antenna, then the strengths of reflected signals measured by the reader are only $-27.3$\,dBm and $-30.3$\,dBm when the tag's antenna is in the short circuit mode and the load match mode, respectively.

%=======================================
\subsubsection{Ambient backscatter WIPT}
Recently, a new backscatter WIPT system, called the ambient backscatter assisted communication, is proposed~\cite{lu2018ambient}. Instead of using the RF wave transmitted from a dedicated reader as the carrier signal and energy medium, the new system harvests and reflects ambient radio waves radiated from TV towers and cellular base stations for wireless communication.

In the ambient backscatter WIPT system, a sender performs the load modulation to change the amplitude of waves reflected from ambient RF. The receiver can be another backscattering node~\cite{liu2013ambient} or a small RF transmitter, e.g., a wireless router~\cite{bharadia2015backfi}. The received data can be decoded by measuring the average strength of backscattering waves arrived in a certain period of time. 

To apply the ambient backscatter WIPT in the real world, one challenge is the irregular carrier problem. Specifically, an ideal carrier signal is a CW wave, whose amplitude and frequency are constant. However, most ambient RF signals are a combination of modulated signals, their amplitude and frequency change with time. As a result, the random fluctuation in amplitudes of the ambient carrier signal may overwhelm the change in intensities of the reflected wave caused by the sender's load modulation. To detect the latter reliably, the transmitter needs to increase the energy per bit by reducing the data rate so that the accumulative energy of reflected waves can have a noticeable difference when different data bits are sent~\cite{liu2013ambient, parks2015turbocharging}. As a consequence, the ambient backscatter WIPT systems usually have a low transmission rate and a short communication range.

%=============================================
\subsection{Far-Field WIPT}
\label{subsec:ArchFarField}
Different from the inductively coupled WIPT and the backscatter WIPT, the far-field WIPT can transmit data actively without any incoming signal as the information carrier. Depending on whether or not the energy harvesting circuit and the communication module share the antenna, far-field WIPT systems can be divided into the following three categories:

\subsubsection{Separated system}
In a separated far-field WIPT system, the energy harvester has a dedicated antenna and therefore can be considered as a battery, which replenishes energy automatically and supplies power to the wireless communication module.

\subsubsection{Co-located system}
\begin{figure}[htb]
\centerline{\includegraphics[width=8.5cm]{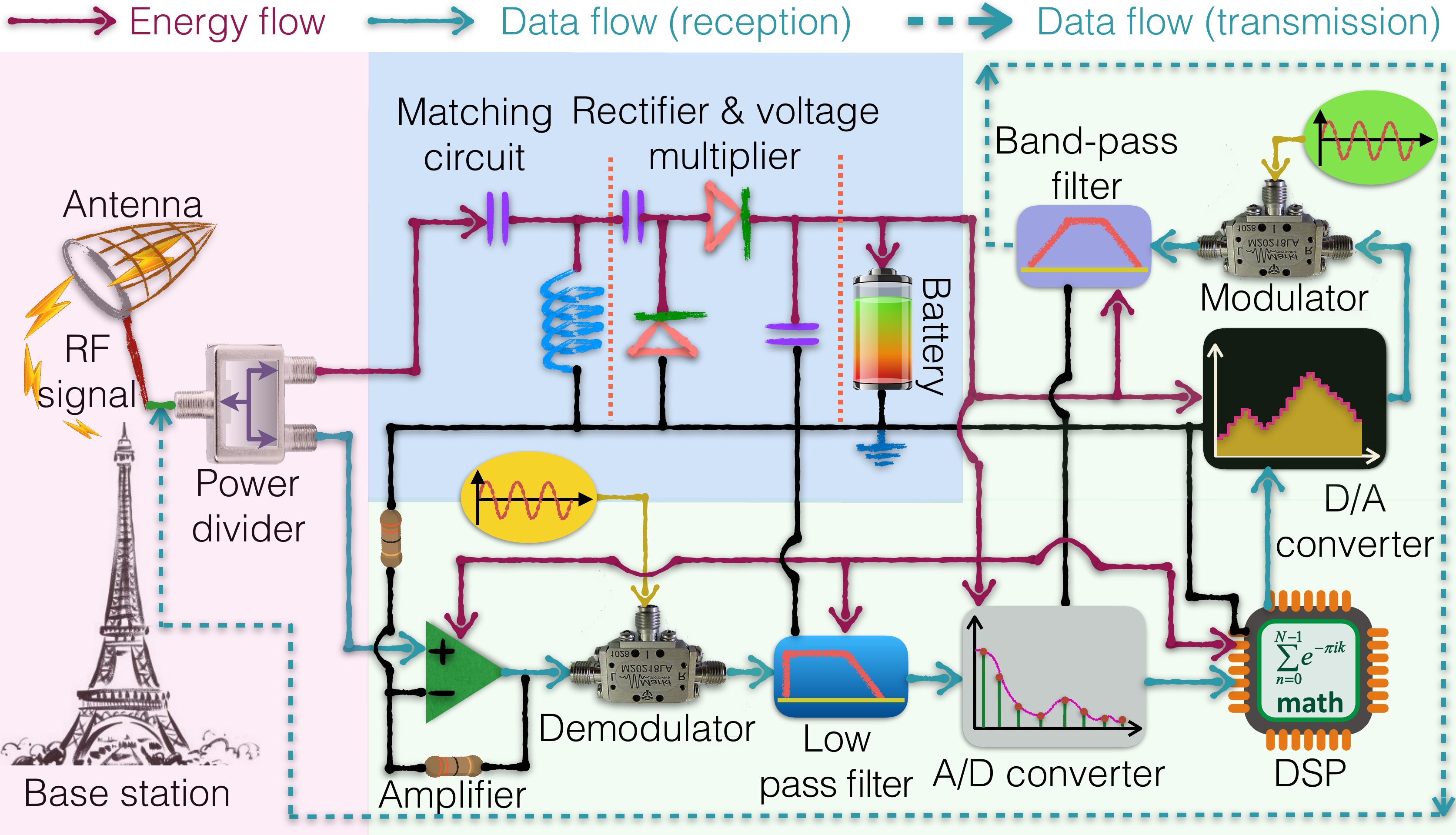}}
  \caption{An example of a far-field WIPT system, where the blue frame and the green frame are the energy harvesting circuit and the communication module, respectively.}\label{fig:structure(d)}
\end{figure}

In a co-located far-field WIPT system, the energy harvester and the communication module share an antenna\cite{zhang2013mimo}, as shown in Fig.~\!\ref{fig:structure(d)}. According to how the energy harvesting process coordinates with wireless communications, a co-located WIPT system has three branches, which are the time-switching, power-splitting, and antenna switching systems. In the time-switching WIPT,  a system can either harvest energy or communicate with other wireless devices, but not at the same time. In the power-splitting WIPT, the incident RF wave is split into two streams by a power divider. The stream went through the communication module is decoded as an information; the stream flowed through the energy harvester is converted into energy to power the whole system. In the antenna switching WIPT, multiple receiving antennas are divided into two groups and each group is arranged to send/receive data or harvest energy.

\subsubsection{Integrated system}
In an integrated WIPT system, the rectifier in the energy harvesting circuit is also used as a low-pass filter to extract the baseband signal from received RF signals~ \cite{zhou2013wireless}. After that, the baseband signal is split into an energy stream and a data stream by a power switcher or a time switcher in order to receive energy and information at the same time.

Compared with a co-located system or an integrated WIPT system, the separated architecture has two advantages. First, it can optimize the energy harvesting efficiency and the data rate simultaneously in a far-field RF environment. To achieve this, separated antennas are necessary since the energy harvest and the wireless communications need antennas with different characteristics~\cite{tran2017rf}. Specifically, to harvest energy efficiently, the antenna should have a ultra-wide frequency band or multiple resonant frequencies to collect energy from a wide range of frequencies, while for wireless communications, an antenna with a moderate bandwidth is preferred in order to maximize the antenna gain on a certain frequency while reducing the wide-band noise~\cite{song2015high, suh2002high}. Secondly, in a separated WIPT system, the energy stream and the data stream can have different frequencies, thereby preventing the RF energy from interfering with information decoding.

In a co-located system, the energy harvester and the communication module share an antenna to reduce the system size. However, sharing the antenna usually requires the energy stream and the information stream to have the same frequency band. Hence, a power divider in the co-located WIPT needs to carefully adjust the distribution ratio between the energy flow and the data flow so that the system can have enough signal-to-noise ratio (SNR) to decode information and sufficient power to drive the system.

\bibliographystyle{IEEEtran}
\bibliography{WIPT}

\end{document}